\documentclass[10pt,conference]{IEEEtran}
\IEEEoverridecommandlockouts

\usepackage{cite}
\usepackage{amsmath,amssymb,amsfonts}
\usepackage{algorithmic}
\usepackage{graphicx}
\usepackage{subcaption}
\usepackage{textcomp}
\usepackage[hyphens]{url}
\usepackage[dvipsnames]{xcolor}
\usepackage{comment}
\usepackage{float}
\def\BibTeX{{\rm B\kern-.05em{\sc i\kern-.025em b}\kern-.08em
    T\kern-.1667em\lower.7ex\hbox{E}\kern-.125emX}}
\usepackage[colorinlistoftodos]{todonotes}

\begin{document}

\pdfoutput=1
\title{Metamorphic Testing for Quality Assurance of Protein Function Prediction Tools\\
}

\author{\IEEEauthorblockN{Morteza Pourreza Shahri*}
\IEEEauthorblockA{\textit{Gianforte School of Computing} \\
\textit{Montana State University}\\
Bozeman, MT, USA \\
mpourrezashahri@montana.edu}
\and
\IEEEauthorblockN{Madhusudan Srinivasan*}
\IEEEauthorblockA{\textit{Gianforte School of Computing} \\
\textit{Montana State University}\\
Bozeman, MT, USA \\
madhusuda.srinivasan@montana.edu}
\thanks{*The authors wish it to be known that Morteza Pourreza Shahri and Madhusudan Srinivasan should be regarded as joint first authors.}
\and
\IEEEauthorblockN{Gillian Reynolds}
\IEEEauthorblockA{\textit{Dept. of Plant Sciences and Pathology} \\ 
\textit{Montana State University}\\
Bozeman, MT, USA \\
gillian.reynolds@student.montana.edu}
\and
\IEEEauthorblockN{Diane Bimczok}
\IEEEauthorblockA{\textit{Dept. of Microbiology and Immunology} \\
\textit{Montana State University}\\
Bozeman, MT, USA \\
diane.bimczok@montana.edu}

\and
\IEEEauthorblockN{Indika Kahanda}
\IEEEauthorblockA{\textit{Gianforte School of Computing} \\
\textit{Montana State University}\\
Bozeman, MT, USA \\
indika.kahanda@montana.edu}

\and
\IEEEauthorblockN{Upulee Kanewala**}
\IEEEauthorblockA{\textit{Gianforte School of Computing} \\
\textit{Montana State University}\\
Bozeman, MT, USA \\
upulee.kanewala@montana.edu}
\thanks{**Corresponding Author.}
}
\maketitle

\title{A Look Back at the Quality of Protein Function Prediction Tools in CAFA}



\begin{abstract}
Proteins are the workhorses of life and gaining insight on their functions is of paramount importance for applications such as drug design. However, the experimental validation of functions of proteins is highly-resource consuming. Therefore, recently, automated protein function prediction (AFP) using machine learning has gained significant interest. Many of these AFP tools are based on supervised learning models trained using existing gold-standard functional annotations, which are known to be incomplete.
The main challenge associated with conducting systematic testing on AFP software is the lack of a test oracle, which determines passing or failing of a test case; unfortunately, due to the incompleteness of gold-standard data, the exact expected outcomes are not well defined for the AFP task. Thus, AFP tools face the \emph{oracle problem}. In this work, we use metamorphic testing (MT) to test nine state-of-the-art AFP tools by defining a set of metamorphic relations (MRs) that apply input transformations to protein sequences. According to our results, we observe that several AFP tools fail all the test cases causing concerns over the quality of their predictions.
\end{abstract}

\begin{IEEEkeywords}
Metamorphic testing, Protein function prediction, Supervised learning
\end{IEEEkeywords}

\section{Introduction}
\subsection{Proteins and their functions}
Proteins are one of the main components of a living body that are important due to various vital functions they perform in living cells. Basically, a cell is alive because of the functions of proteins. While our genes encode protein sequences, proteins determine all other aspects of cell function including metabolism, structure, transport, signaling, immune defense, cell division and cell death. Disease processes associated with hereditary genetic defects ultimately are due to dysfunctions in the proteins that the genes encode. 

Various forms of Alzheimers, Huntingdons, Parkinsons, cystic fibrosis and hemophilia are all well-known examples of protein misproduction caused by errors in the underlying genetic code~\cite{cutting2015cystic, ferreira2017updated,karch2014alzheimer, prasad2018hemophilia, lee2015identification}. Lesser-known examples include errors in the BRCA1 and BRCA2 genes, which are known to increase a persons risk of developing breast cancer, and errors in the code of Msh2, which increases a risk of developing colon and endometrial cancers~\cite{trujillano2015next, sehgal2014lynch}. 
Whilst all of the aforementioned diseases are vastly different in their epidemiological background, one element they all have in common is a disruption of a proteins ability to correctly perform its function.

Gene Ontology (GO) is a framework used for describing protein functions~\cite{ashburner2000gene}. 
Gene Ontology is composed of different classes (or terms), each of which demonstrates a single function, and the hierarchical relations between the classes. Within the GO term hierarchy, \emph{child} terms are more specialized than their \emph{parent} terms, e.g., \emph{tyrosine metabolic process} is a child term of \emph{metabolic process}.  In addition, GO relations can be \textit{is-a} relations, \textit{part-of} relations, etc. For example, the protein BRCA1\_HUMAN has a list of functions such as \textit{androgen receptor binding}~(GO:0050681), \textit{damaged DNA binding}~(GO:0003684), etc.

Gene Ontology is composed of three sub-ontologies: the \emph{molecular function (MF)} ontology, which describes various molecular activities, the \emph{biological process (BP)} ontology, which describes various processes that a protein may be involved with and the \emph{cellular component (CC)} ontology, which describes the localization of proteins.
The official Gene Ontology website\footnote{http://www.geneontology.org/} maintains not only the ontology but the annotations using the ontology (the gold-standard functional annotations) for a large collection of proteins from many different organisms~\cite{ashburner2000gene}. Many of these annotations are experimentally validated through wet-lab assays. These annotations follow the ``true path rule" which means annotations to a certain term imply annotations to all of its ancestors~\cite{rhee2008use}.


However, how biologists identify such function has been drastically altered over the last decade, thanks to the next generation sequencing revolution. Following the completion of the human genome project, DNA sequencing technology has developed at such a rate that it far surpassed Moores law~\cite{gullapalli2012next}. The most striking example of this is the cost of sequencing a human genome. 15 years ago, the completion of the human genome project was announced. This project was a large international collaboration which took 13 years and \$2.7 billion to complete~\cite{national2010human}. In a clinical setting today, the cost of whole genome sequencing has been reported to be approximately \$1,906-\$24,810 and it could be sequenced and assembled in a matter of weeks~\cite{schwarze2018whole}, and based on the genetic code, protein sequences can be directly inferred from gene sequences.  The exponential growth of available gene and protein sequence data presents a whole new suite of challenges to today's biologists rendering the gold-standard Gene Ontology annotations incomplete. It is reported that only a small percentage of known proteins have experimentally validated annotations, while many among them are considered incomplete\cite{gene2018gene}. 
This has highlighted the need for high-throughput approaches for functional annotation and has consequently fostered collaborations between a range of disciplines, most notable of which is computer science. 

\subsection{Automated Protein Function Prediction (AFP)}
Through the development of numerous algorithms and tools, collaborations with bioinformaticians and computational biologists have altered the way and speed in which biologists can make sense of the deluge of genomic data. One area that has benefited significantly from such developments is automated protein function prediction (AFP). As mentioned above, previous routes of ascertaining protein function required extensive wet-lab investigations, often only focusing on one protein at a time and could be considered low throughput~\cite{shehu2016survey}. Whilst such experiments are still required for validation, computational protein function prediction tools have significantly changed the way biologists conduct protein function investigations. These high throughput approaches have been essential for modeling the impact that errors in the genetic code have upon the function of proteins and how this impacts the health of an organism. 



Automated function prediction tools typically take a protein sequence as their input and output a set of predictive GO terms corresponding to their  functional categories. These protein sequences are stored in the text-based FASTA\footnote{\url{https://blast.ncbi.nlm.nih.gov/Blast.cgi?CMD=Web\&PAGE\_TYPE=BlastDocs\&DOC\_TYPE=BlastHelp}} format where the protein sequence is preceded by a description line, identified by the ``\textgreater" symbol.



These tools typically make their predictions using various techniques such as sequence matching that employ the sequence alignment to extract the functions of similar proteins, protein structure-based methods, genomic context-based methods, phylogenomics-based methods, protein-protein interaction-based methods, data integration methods, and text mining-based methods~\cite{shehu2016survey}.

Sequence-based methods match a large collection of sequences with a target protein sequence, and using this comparison they determine whether the sequences under comparison share a common ancestor. One subgroup of sequence-based methods is the methods that fall in the same category but do not directly predict functions of proteins, but they provide information about protein sequences by extracting features which can be used by other machine learning methods~\cite{pfp,cons,argot,pannzer}. 

Protein structure-based methods try to find a level of similarity using two given protein structures which provides the transfer of functional annotations between proteins, and the similarity can be detected using the entire structure or only a part of the structures~\cite{molloy2014exploring,braberg2012salign}.

Genomic context-based methods rely on the knowledge that the location of the gene which is encoding a query protein is prominent information that can be used for function prediction~\cite{korbel2004analysis,ferrer2010systematic}. Evolutionary relationships are also exploited between organisms to find functional similarities between genes in phylogenomics-based methods~\cite{gorbi,cbrg}.

Interaction-based methods utilize protein-protein interaction (PPI) networks in which PPI data is represented as vertices (proteins) and edges (direct bindings). These interactions can be utilized to find functional relationships, and to achieve this goal, graph-theoretic methods and algorithms can be employed to predict functions of proteins~\cite{profun,wang2015explore}.

The data integration-based methods are mostly based on machine learning in which features generated from different biological sources are combined and used for training a machine learning model~\cite{inga,wass2012combfunc}.

Text mining-based methods have been employed for the analysis of biomedical literature for the problem of protein function prediction with the idea that the large amount of information in the literature can link proteins with each other. Therefore, they can be utilized to increase the size of labeled data for the task of training and evaluation~\cite{evex,raychaudhuri2002associating}.


Critical Assessment of protein Function Annotation~(CAFA) is a community-wide large-scale evaluation of AFP tools organized by the Function Special Interest Group\footnote{https://biofunctionprediction.org/}.
At the time of writing this paper, CAFA2 was the latest challenge where its results of evaluation were publicly available~\cite{cafa2}. Many tools were presented in CAFA2, and they were evaluated using different criteria such as macro-AUROC, F-max, and $S_{min}$~\cite{cafa2}.

\subsection{Quality Assurance of AFP tools}
Despite the plethora of AFP tools and comprehensive CAFA evaluation results, selecting a tool from this list of top performing AFP tools to perform experiments or research would be very challenging as described below. One way to select a tool is randomly picking a few tools, feeding well-known protein sequences into the tools, and comparing the outputs with the experimentally validated GO terms, which are the results of a physical characterization of a gene product that has supported its association with the GO term\footnote{http://www.geneontology.org/page/guide-go-evidence-codes}.

\begin{figure*}[t]
\centering
\includegraphics[width=0.7\linewidth]{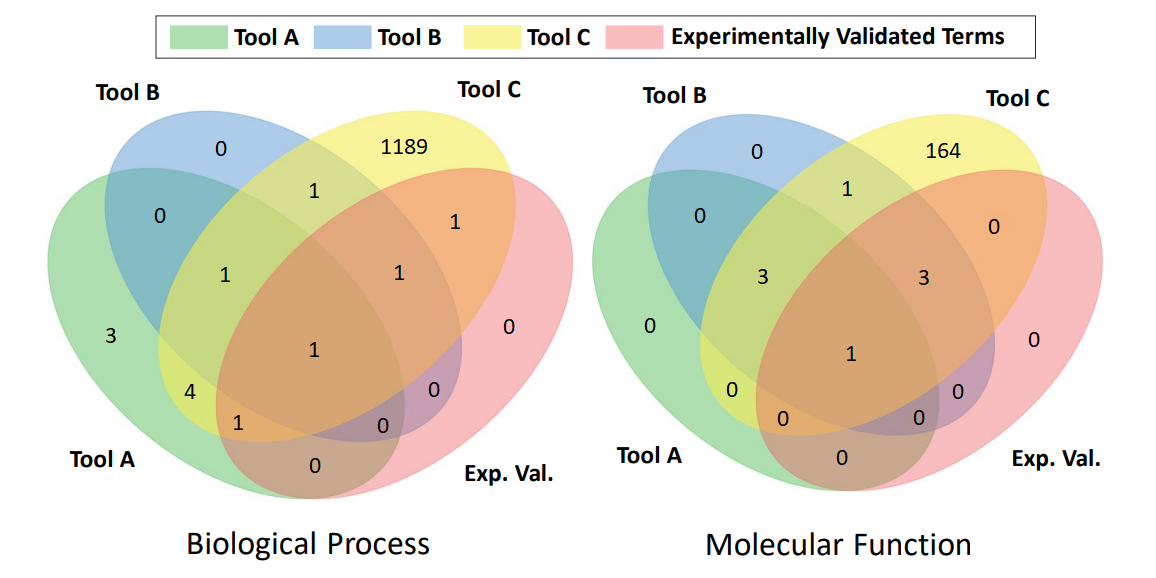}
\caption{Distribution of output GO terms for the protein TYRO on Biological Process and Molecular Function ontologies}
\label{fig:venn}
\end{figure*}

However, using a few tools and sequences in the above specified manner, users would observe that each tool provides different set of output GO terms, and only a few terms would be in common with the experimentally validated terms. Fig.~\ref{fig:venn} shows the distribution of predicted GO terms using three randomly selected top performing CAFA2 tools in comparison with the corresponding experimentally validated terms of the well-known protein \textit{Tyrosinase}, an enzyme that hydroxylates tyrosine as the first step in melanin synthesis\footnote{Diagrams generated by https://github.com/tctianchi/pyvenn}. The distribution of GO terms in both the  Molecular Function and Biological Process ontologies shows that only one of the predicted GO terms is in common between the three tools and the experimentally validated terms. 
The next important observation on Biological Process ontology is that one of the tools returns 1199 GO terms, whereas another tool outputs only four GO terms for the same protein. 
Moreover, as mentioned above, the experimentally validated terms set is incomplete. Therefore, these observations show why it would be challenging for a  biologist to select a tool for their research, as well as for a developer to test the tools that they develop.

Yet, protein function prediction tools form an essential part of the vast majority of protein function investigations. Designed to complement rather than replace experimental analysis, these tools are often employed to direct the focus on experimental investigations. Failure of the tools to perform accurately could lead to lengthy, expensive and ultimately fruitless experimental investigations. Thus, it is essential to develop cost effective approaches for systematically testing AFP tools.

In this study, we apply metamorphic testing~(MT) for the quality assurance of AFP tools. We develop novel metamorphic relations (MRs) using transformations to protein sequences that are typically used as inputs to AFP tools. We use these MRs to test nine top performing AFP tools from CAFA2. The results of this study show that several AFP tools fail all the test cases and only at most two tools pass all the test cases. Therefore this study has implications for both the the developers and the users of these AFP tools.



\section{Metamorphic Testing}
In complex systems, such as AFP tools, it is practically difficult to determine whether the output provided by the system for a given input is correct. This is known as the \emph{oracle problem}~\cite{Weyuker01111982}. MT can be used to test programs that face the oracle problem~\cite{chen1998metamorphic}. The MT process involves deriving MRs and generating test cases based on those MRs. A MR is a relation derived from the specification of the program under test and specifies how the output would change according to a specific change made to the input. \emph{Source test cases} are typically derived using a traditional test case generation approach such as random test generation. Typically, the \emph{Follow-up test cases} are derived by applying the transformations specified in the MR to the source test case and/or source outputs~\cite{chen2018metamorphic}.
Then, the source and follow-up test cases are executed and outputs of these test cases are used to verify whether MR was violated or not. The violation of a MR indicates faults in the program. 

\begin{figure}[ht]
\includegraphics[width=\linewidth]{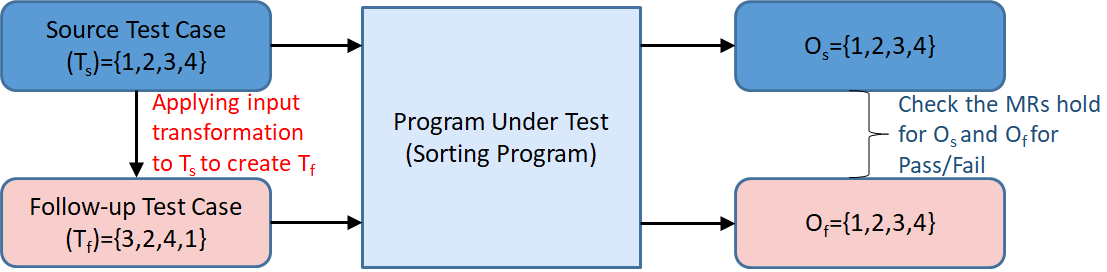}
\caption{MT example for sorting program}
\label{fig:sorting}
\end{figure}

Fig.~\ref{fig:sorting} shows how MT is applied to a sorting program. This sorting program arranges a random set of numbers provided as input in the ascending order. A MR derived for the sorting program states that \emph{when the original set of numbers are shuffled and used as an input to the program, the output must be equal to the original output.}
In order to conduct MT on the sorting program using this MR, the source test case can be created by generating a set of random numbers and the follow-up test case can be created by shuffling the source test case. A fault is detected in the sorting program if the outputs from the source and follow-up test cases are not equal as defined in the MR~\cite{chen2018metamorphic}.

\section{Metamorphic Testing for AFP Tools}
The first step of applying MT for testing a given program is defining MRs. The most commonly used approach to define MRs is looking at the changes that you can make to the input and whether those changes would cause predictable changes in the output. However, defining MRs for AFP tools should be done with caution because, we cannot make random changes to the input sequences since the input sequence represents a specific protein and such changes would cause the sequence to loose its meaning. Any changes that we make to the input sequence should be made based on relevant biological knowledge as we discuss below. MRs are designed to achieve better understanding of the software .

We define a MR using the \emph{canonical sequences} and their \emph{variants}. A canonical sequence is defined as the ``standard'' sequence, generally based on its prevalence in the population and its similarity to \emph{orthologous sequences} in other species. The term \emph{orthologous sequence} is used in biology to refer to similar genetic sequences that are found in other species.
Generally speaking, these orthologous genetic sequences are thought to maintain a similar function across the species it can be found in.
All other sequences are hence considered variants of the canonical sequence. These sequence variants include genetic polymorphisms, disease-associated mutations and RNA editing events such as alternative splicing. Both the canonical sequence and the variants are generally listed under one single entry in the UniProt/Swissprot databases\footnote{https://www.uniprot.org/} (which are the primary knowledge bases on proteins).

For testing AFP tools, we define a MR in the broad sense that says \emph{there should be a change in the output GO terms between the canonical proteins and their corresponding well-studied variants}. 
Note that this assumption does not always hold true for all proteins, but we have carefully chosen only the protein examples that satisfy our MR. Thus, this MR imposes restrictions on the source test cases that can be used. More specific instances of this MR can be created by observing the characteristics of different variants. For instance, if the source test case is the canonical sequence of the protein \textit{Tyrosinase}, and the follow-up test case is a disease variant of this protein which causes \textit{Albinism (OCA1A)}, biological knowledge entails that the output GO terms of the canonical sequence and the disease variant must be different. 

We also note that the \emph{change} in the output mentioned above is measured using the \emph{set} difference. In other words, if the set of GO terms for the variant sequence is different from the set of GO terms for the canonical sequence, it is considered a change. In this setting, a GO term is only a match (i.e. equal) to that term itself, but not to any of its ancestors and\ or descendants. This interpretation is consistent with the CAFA evaluation setup in which tools are penalized for predicting a GO term that is an ancestor or a descendant of the gold standard GO term annotation\cite{cafa2}.
\begin{figure}[ht]
\includegraphics[width=\linewidth]{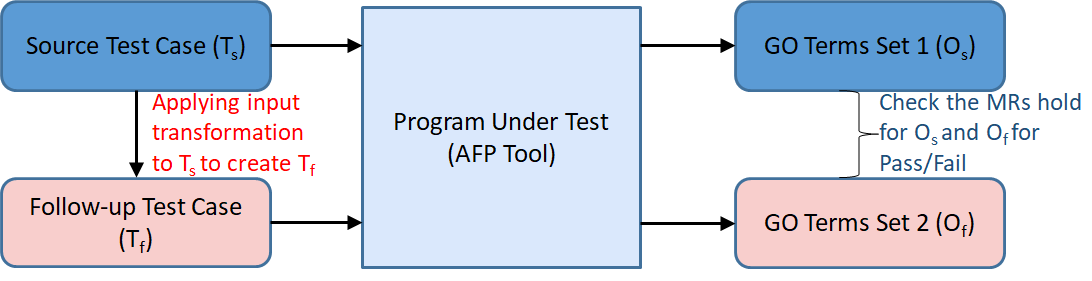}
\caption{Architecture of the Metamorphic Testing system on AFP tools}
\label{fig:mt_afp}
\end{figure}

We use this MR to conduct MT on a given AFP tool by performing the following steps (Fig. \ref{fig:mt_afp} depicts this process):
\begin{enumerate}
\item Running the program with the canonical sequence (i.e. source test case) and getting $O_s$ as the output. The source test case is a FASTA sequence and $O_s$ will be a set of GO terms. 
\item Generating a follow-up test case using the source case, and executing the program with the follow-up test case and getting $O_f$. The follow-up test case is also a FASTA sequence derived from a known variation of the source test case, and the output is a set of GO terms as well.
\item Checking whether the MR defined above holds for $O_s$ and $O_f$. In this example, the MR holds if there is a change in the list of output GO terms, i.e. additions, deletions, etc. If the expected change is satisfied, it will be a \textit{pass}, otherwise, it will be a \textit{fail}.
\end{enumerate}


\subsection{AFP Tool Selection Criteria}
\begin{figure*}
\centering
\includegraphics[width=0.7\linewidth]{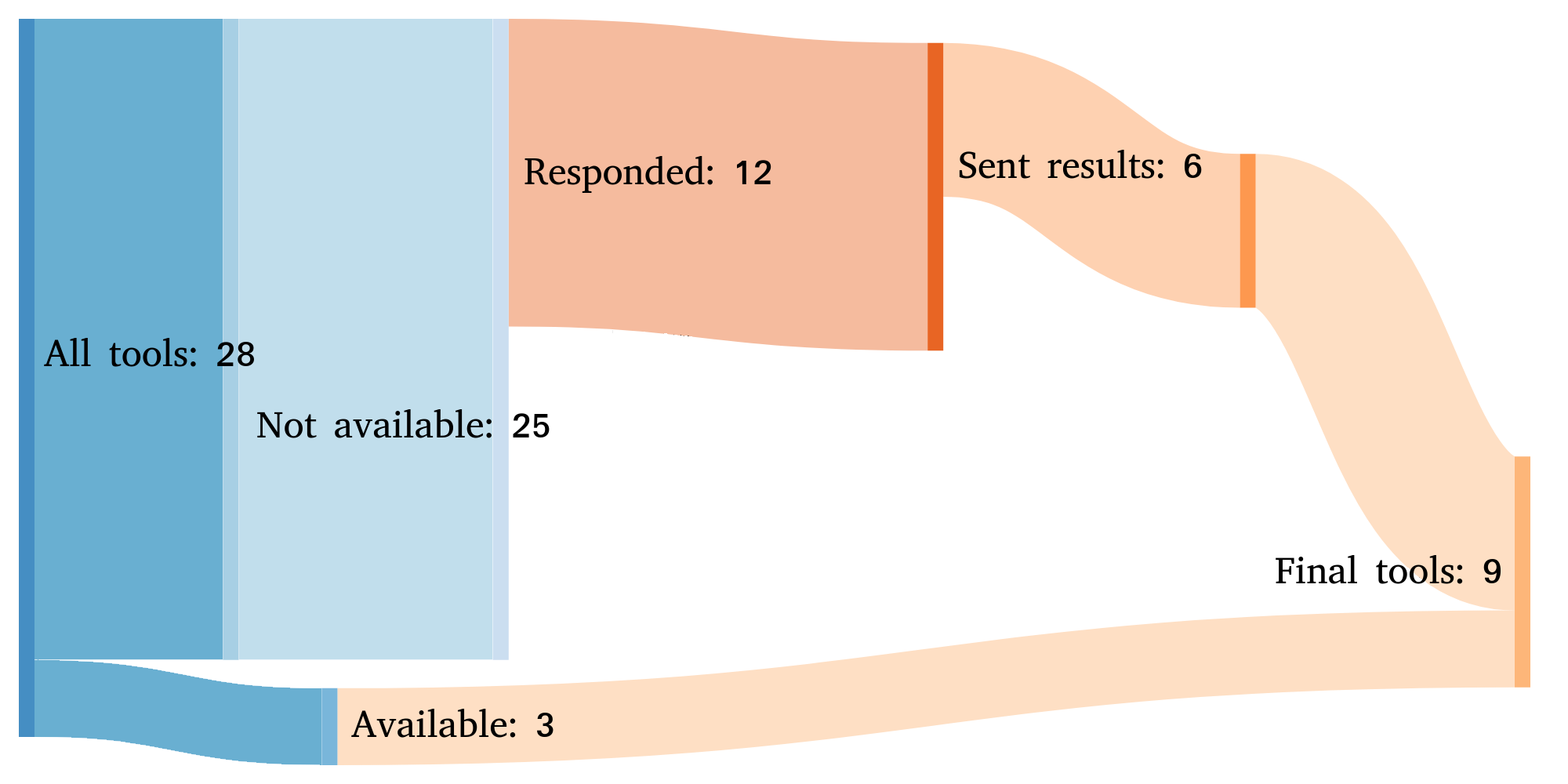}
\caption{The flow of selection of the tools}
\label{fig:flow}
\end{figure*}

In order to identify a suitable set of AFP tools to apply MT, we started with the 28 top-performing tools from the CAFA2 challenge. From these 28 tools, most are not publicly available, and some are very hard to setup and run. So, we selected tools that can be set-up for execution by spending a maximum of thirty minutes by a graduate student. At the time of this investigation, only three tools were publicly available and/or worked as advertised. As we wanted to perform the experiments on as many tools as possible, we contacted authors of the remaining 25 tools, requesting them to feed the sequences used as source and follow-up test cases to their tools and provide us with the outputs. Twelve authors responded positively, and 6 out of 12 authors sent us the outputs. Thus, in our evaluation we used the following nine tools: EVEX~\cite{evex}, PFP~\cite{pfp}, CONS~\cite{cons}, GORBI~\cite{gorbi}, CBRG~\cite{cbrg}, ProFun~\cite{profun}, PANNZER~\cite{pannzer}, Argot2~\cite{argot}, and INGA~\cite{inga}. Fig.~\ref{fig:flow} depicts above mentioned work-flow of selecting the nine tools used in the evaluation.


\subsection{Source and Follow-up Test Cases}


\begin{table}[h]
 \caption{The List of Selected Proteins}
\label{tab:proteins_list}
\centering
 \begin{tabular}{| c |p{4cm} |c |} 
 \hline
  & Protein Name & UniProt Id \\
 \hline
 TYRO\_Human & Tyrosinase & P14679\\ 
 \hline
 IL2RG\_Human & Cytokine receptor common subunit gamma & P31785\\ 
 \hline
 TLR4\_Human & Toll-like receptor 4 & O00206\\ 
 \hline
\end{tabular} 
\end{table}


We used 18 sequences from three carefully selected well known proteins as source and follow-up test cases for testing the AFP tools using the previously defined MR. These proteins are shown in Table~\ref{tab:proteins_list}. Tyrosinase (TYRO) and interleukin-2 receptor gamma (IL2RG, also termed the common gamma chain) were selected, because they have well characterized and highly defined functions. Importantly, 
point mutations (modification to a single location in the sequence) in the TYRO and the IL2RG 
cause a loss of a particular protein function that directly results in a clinical disease, i.e., oculocutaneous albinism and severe combined immunodeficiency, respectively. Toll-like receptor 4 (TLR4) likewise is a very well characterized innate immune receptor that mediates activation of pro-inflammatory signaling pathways upon binding of bacterial material. 
With respect the three selected proteins, changes in protein functions due to an altered amino acid sequence are expected to result in changes in GO terms for Molecular Function and Biological Process. We do not anticipate alterations in Cellular Component ontology.  


These three proteins have a large numbers of variants associated with each of them. For example, TYRO has 99 variants involved in oculocutaneous albinism type A (OCA1A) alone, therefore, it not feasible and cost effective to execute the tools using all these variants.

We selected the number of variants to execute for each protein proportional to its sequence length. Thus, for TYRO more variants would be selected for execution compared to IL2RG since TYRO sequence is longer than IL2RG, i.e. we selected seven variants for TYRO, four variants for IL2RG, and four variants for TLR4. 
In the next step, we divide the sequence into equal segments proportional to the number of selected variants, i.e. for TYRO we divide the sequence into seven equal segments. Next, from each segment we pick the variant with the largest number of associated  publications, which provides more experimental evidence for the existence of the variant.
Eventually, the sequences consist of the canonical, i.e. standard, sequence and the sequences of variants as follows:

\begin{itemize}
\item TYRO\_HUMAN: (Canonical sequence + 7 disease variants)
\item IL2RG\_HUMAN: (Canonical sequence + 4 disease variants)
\item TLR4\_HUMAN: (Canonical sequence + 2 splice variants + 2 natural variants)

\begin{table}
 \caption{Variants Selection Criteria}
\label{tab:variant_selection_criteria}
\centering
 \begin{tabular}{| c|c |c |c |} 
 \hline
 Protein & Identifier & Position & Change \\
 \hline\hline
 TYRO\_Human & P14679 &  & Canonical\\ 
 \hline
 TYRO Variant 1 & VAR\_007652 &  47 & G - D \\ 
  \hline
 TYRO Variant 2 & VAR\_007658 & 81  & P - L \\ 
  \hline
 TYRO Variant 3 & VAR\_007667 &  217 & R - Q \\ 
  \hline
 TYRO Variant 4 & VAR\_007671 & 299 & R - H \\ 
  \hline
 TYRO Variant 5 & VAR\_007680 & 373  & T - K \\ 
  \hline
 TYRO Variant 6 & VAR\_007690 & 419 & G - R \\ 
  \hline
 TYRO Variant 7 & VAR\_007692 & 446  & G - S \\ 
  \hline
 IL2RG\_Human & P31785 &  & Canonical \\ 
  \hline
 IL2RG Variant 1 & VAR\_002668 & 39 & D - N \\ 
  \hline
 IL2RG Variant 2 & VAR\_002681 &  153 & I - N \\ 
  \hline
 IL2RG Variant 3 & VAR\_002690 &  226 & R - C \\ 
  \hline
 IL2RG Variant 4 & VAR\_002701 &  285 & R - Q \\
 \hline
 TLR4\_Human & O00206 &  & Canonical \\ 
 \hline
 TLR4 Natural 1 &  & 526 & N - A \\ 
 \hline
 TLR4 Natural 2 &  & 711 & D - K \\ 
 \hline
 TLR4 Splice 1 & O00206-2 &  & Isoform 2 \\ 
 \hline
 TLR4 Splice 2 & O00206-3 &  & Isoform 3 \\ 
 \hline
\end{tabular} 
\end{table}

\end{itemize}

Table~\ref{tab:variant_selection_criteria} demonstrates the exact changes in the canonical sequences of proteins and their corresponding positions in the sequence. For each protein, we use the canonical sequence as the source test case and each of the variant sequences as the follow-up test case. Therefore, we have 15 pairs of source and follow-up test cases.

\subsection{Test Execution}
The next step in applying MT to AFP tools is to feed the test cases into the tools, and checking whether the MRs hold for each execution. Therefore, we have nine tools and 15 pairs of source and follow-up test cases. 

For each pair of source and follow-up test cases, we store the output GO terms for the Molecular Function and Biological Process ontologies separately, and compare the GO terms of $O_s$ and $O_f$, and report the results of different ontologies separately.

\section{Results}
\begin{figure}
    \centering
    \begin{subfigure}[b]{0.45\textwidth}
        \includegraphics[width=\textwidth]{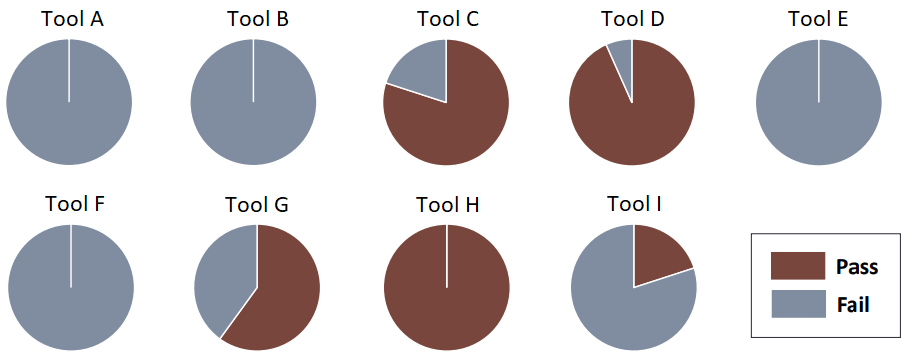}
        \caption{Molecular function ontology}
        \label{fig:res_mf}
    \end{subfigure}
    \\ 
    \begin{subfigure}[b]{0.45\textwidth}
        \includegraphics[width=\textwidth]{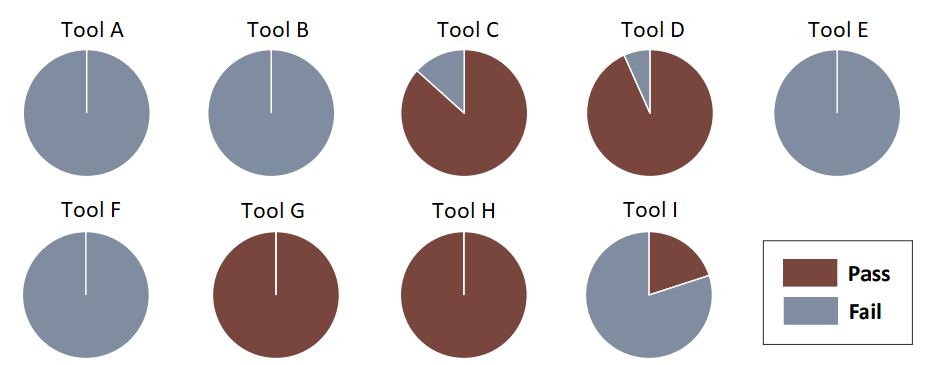}
        \caption{Biological process ontology}
        \label{fig:res_bp}
    \end{subfigure}
    \caption{Overall test results}
\end{figure}

Figs. \ref{fig:res_mf} and \ref{fig:res_bp} show the results of executing the tools with 15 test case pairs on Molecular Function ontology and Biological Process ontologies, respectively. Each pie chart shows the number of passes and fails of the 15 test case pairs for a given tool. As shown in Fig.~\ref{fig:res_mf}, only tool H passes all the test cases. Four out of the nine tools fail all the test cases. This phenomenon can happen if the tools are not designed to detect variations in the protein sequence. 
The rest of the tools have a mix of passes and fails.

We executed the same 15 test case pairs (also known as metamorphic Group of Inputs~\cite{zhou2018metamorphic}) on the Biological Process ontology as well. As shown in Fig.~\ref{fig:res_bp}, two tools, G and H passed all the test cases. Four tools, A, B, E and F failed all the test cases. Interestingly, these are the same tools that failed all the test cases for the Molecular Function ontology. This further validates our hypothesis that these four tools are not designed to detect variations in protein sequences.  

\begin{figure}
    \centering
    \begin{subfigure}[b]{0.4\textwidth}
        \includegraphics[width=\textwidth]{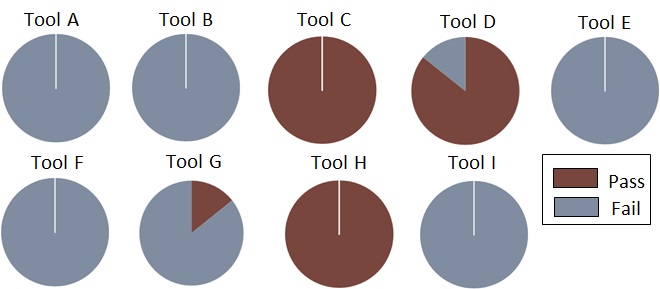}
        \caption{TYRO}
        \label{fig:TYRO_mol_tools}
    \end{subfigure}
    \\ 
    \begin{subfigure}[b]{0.4\textwidth}
        \includegraphics[width=\textwidth]{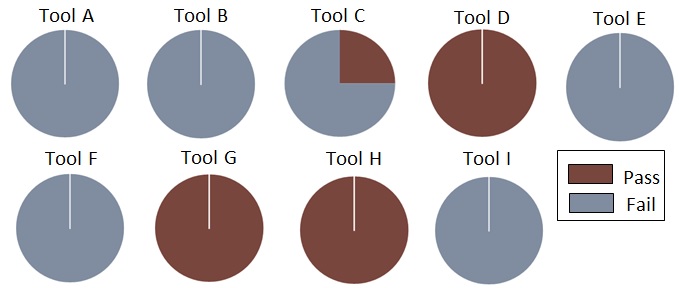}
        \caption{ILR2G}
        \label{fig:IL2RG_mol_tools}
    \end{subfigure}
    \\
    \begin{subfigure}[b]{0.4\textwidth}
        \includegraphics[width=\textwidth]{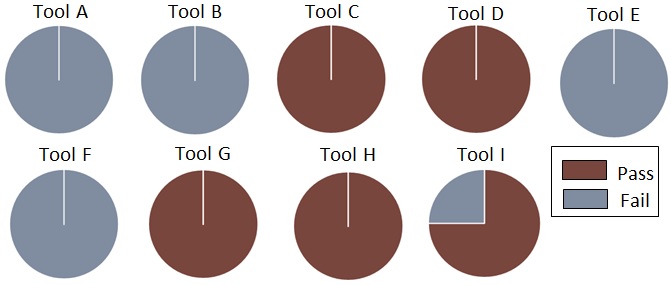}
        \caption{TLR4}
        \label{fig:TLR4_mol_tools}
    \end{subfigure}
    \caption{Test results at the protein level for the molecular function ontology}
\end{figure}

Next we analyze the test results at the individual protein sequence levels for the two ontologies. Figs.~\ref{fig:TYRO_mol_tools}, ~\ref{fig:IL2RG_mol_tools}, and~\ref{fig:TLR4_mol_tools} show the pie charts of the test results for the Molecular Function ontology for individual protein sequences TYRO, IL2RG and TLR4, respectively. As expected, tool H passed all the test cases. Further, tools D and G passed all the test cases for IL2RG and TLR4. However the performance of tool G on TYRO is not satisfactory as shown in Fig.~\ref{fig:TYRO_mol_tools}. Thus, in addition to tool H, tool D could be another option to use when working with the Molecular Function ontology.

\begin{figure}
    \centering
    \begin{subfigure}[b]{0.4\textwidth}
        \includegraphics[width=\textwidth]{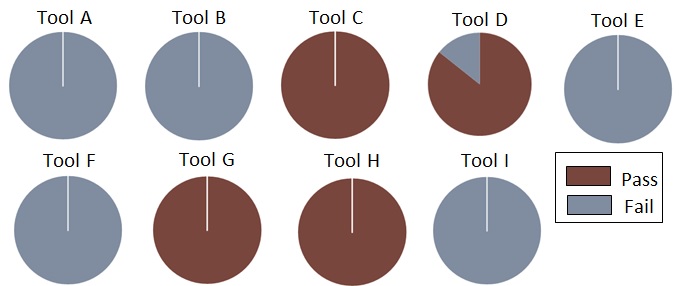}
        \caption{TYRO}
        \label{fig:Tyro_bio_tools}
    \end{subfigure}
    \\ 
    \begin{subfigure}[b]{0.4\textwidth}
        \includegraphics[width=\textwidth]{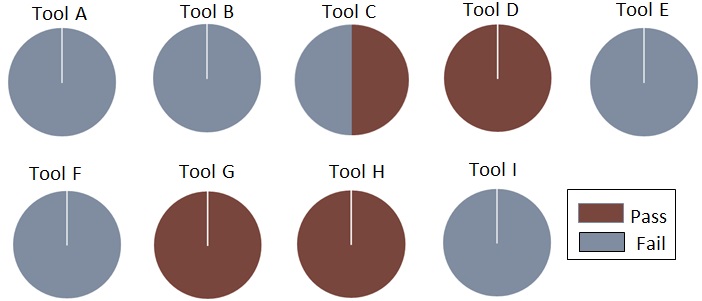}
        \caption{ILR2G}
        \label{fig:IL2RG_bio_tools}
    \end{subfigure}
    \\
    \begin{subfigure}[b]{0.4\textwidth}
        \includegraphics[width=\textwidth]{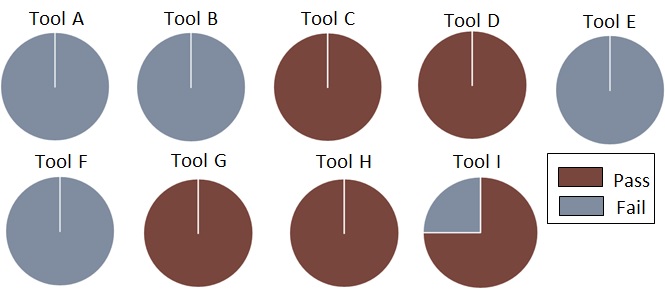}
        \caption{TLR4}
        \label{fig:TLR4_bio_tools}
    \end{subfigure}
    \caption{Test results at the protein level for the biological process ontology}
\end{figure}

Similarly, Figs.~\ref{fig:Tyro_bio_tools},~\ref{fig:IL2RG_bio_tools}, and~\ref{fig:TLR4_bio_tools} show the test results for the Biological Process ontology for the three protein sequences. As expected, tools G and H passed all the test cases for this ontology. Besides them, tool D passed all the test cases for IL2RG and TLR4 and also performed satisfactorily on TYRO. Thus, tool D can be another option to use when working with the Molecular Function ontology.

\begin{figure}
    \centering
    \begin{subfigure}[b]{0.45\textwidth}
        \includegraphics[width=\textwidth]{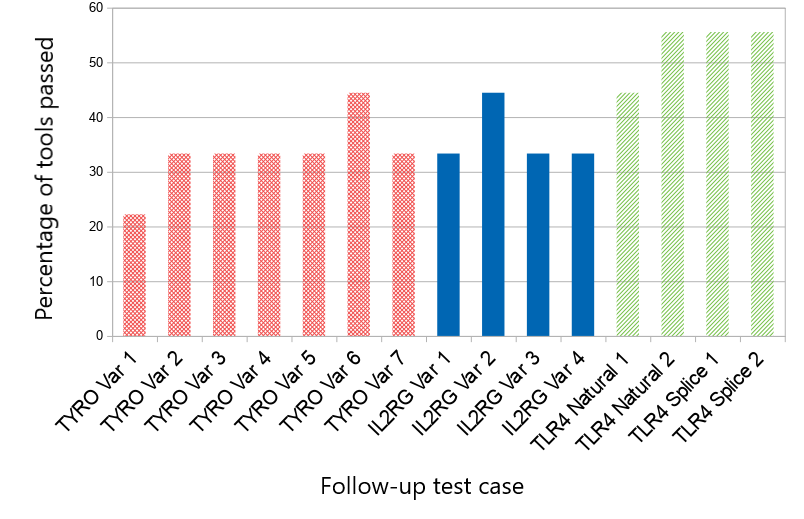}
        \caption{Molecular function ontology}
        \label{fig:pass_percentage_mf}
    \end{subfigure}
    \\ 
    \begin{subfigure}[b]{0.45\textwidth}
        \includegraphics[width=\textwidth]{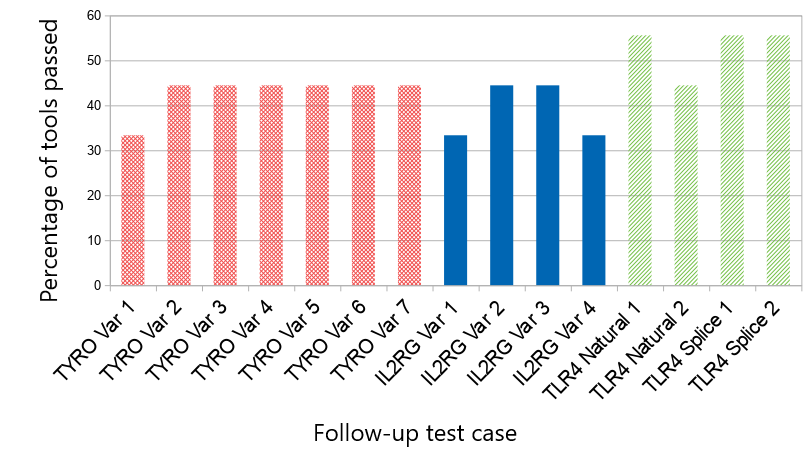}
        \caption{Biological process ontology}
        \label{fig:pass_percentage_bp}
    \end{subfigure}
    \caption{Percentage of tools passed for each source and follow-up test case pair}
\end{figure}

Next, we investigate whether making predictions on certain protein sequences is harder than the others.  Figs.~\ref{fig:pass_percentage_mf} and~\ref{fig:pass_percentage_bp} show the percentage of tools that successfully predicted the changes for each test case pair (named with the corresponding variant used in the follow-up test case) for the Molecular Function ontology and the Biological Process ontology, respectively. We observe that for both ontologies, higher percentage of tools passed the test cases for the variants of TLR4 compared to the other two protein sequences. This observation may suggest that the AFP tools predict the functions of natural and splice variants better than disease variants. More test executions are needed to confirm this hypothesis.

\section{Related Work}
Srinivasan et al. worked on applying MT to LingPipe, a tool for processing text using computational linguistics, which is often used in bioinformatics for bio-entity recognition from biomedical literature ~\cite{srinivasan2018quality}. 
The authors proposed 10 novel MRs and the fault detection effectiveness of each of the MRs was evaluated using mutation testing. Lundgren et al. examined the effectiveness of MT for testing a genome alignment tool BBMap~\cite{lundgren2016experiences}. The experiment results showed that MT is effective in identifying subtle faults compared to pseudo-oracles. Ramanathan et al. used MT to test a workflow of epidemiological models~\cite{ramanathan2012verification}. They showed that MT can be useful when mathematical models fail. Pullum and Ozmen showed that MT could be effective in testing epidemiological models~\cite{pullum2012early}. They used a differential equation and agent-based models for generating MR-transformed parameter values. Chen et al. used MT to test two open-source bioinformatics programs~\cite{chen2009innovative}. The first program GNLab, a tool for large-scale analysis and simulation of gene regulatory networks. The second tool SeqMap deals with mapping a short sequence that reads with a reference genome. The mutants were generated for the GNLab and SeqMap tools. The MRs had different fault finding abilities and the mutants were violated by at least one MR. The MRs related to the change in the nodes in GNLab network were less effective than the other MRs.
Eleni et al. conducted metamorphic testing on three commonly used NGS (Next Generation Sequencing) short-read alignment programs: BWA, Bowtie, and Bowtie2~\cite{giannoulatou2014verification}. The results show that the MR created by permuting reads and addition of reads does not hold for BWA. Also MRs that reverse complement and extend the read fail on both bowtie and BWA.  


\section{Conclusions and Future Work}
In this study, we applied MT for testing nine AFP tools. We use the biological knowledge about proteins and their variants to define an MR that specifies that there should be a change in the predicted GO terms between the canonical protein sequence and their variants. We used this MR to create source and follow-up test cases using carefully selected protein examples such as disease variants. 

Our results indicate that several tools do not pass any of the test cases that we used in this study. This is surprising considering the fact that all of these tools (except one) are among the top performing tools in CAFA2. The only tool for which this failure is to be expected is tool E as it appears this tool was designed specifically for the use of bacterial and archael genomes only (since we used only human data for testing). 

However, it is also possible that these tools are not designed to handle variants. If that is the case, such limitations should be documented such that users of the tools are aware of the limitations~\cite{zhou2018metamorphic}. This would ensure that the biologists, who are the intended primary users of these tools, use them for the appropriate use cases. This is of utmost importance due to the fact that predictions from many of these tools will be used to guide the wet-lab experiments. Predictions lacking in quality could alter the research directions, rendering loss of resources and most importantly can have a significant impact on healthcare applications. 

In the future, we plan to develop more MRs for this domain which uses biological knowledge that will incorporate different orthogonal aspects of AFP such as functional characteristics unique to different species. Further, we plan to develop an expanded test suite by exploring MRs for cellular component sub-ontology and Human Phenotype ontology (which is another structured vocabulary for describing phenotype abnormalities associated with human diseases), and by employing different types of protein examples. We will also work with the AFP community to increase the number of tools to be tested. Eventually, we will develop a testing framework which is readily available for the users and developers in this domain.

\section*{Acknowledgments}
We thank the PIs and teams that provided us with the results of their tools including EVEX~(PI: Filip Ginter), Orengo-FunFams~(PI: Christine Orengo), SIFTER~(PI: Steven Brenner), Jones-UCL~(PI: David Jones), Paccanaro Lab~(PI: Alberto Paccanaro), ProFun~(PI: Jianlin Cheng), PANDA~(PI: Zheng Wang), CBRG and GORBI~(PI: Christophe Dessimoz), FANN-GO~(PI: Predrag Radivojac), CONS and PFP~(PI: Daisuke Kihara). We also would like to thank Dr. Iddo Friedberg and Dr. Rachael Huntley for insightful discussions.

\bibliographystyle{IEEEtran}
\end{document}